\documentclass[aps,showpacs,superscriptaddress,twocolumn]{revtex4}%
\usepackage{mathrsfs}
\usepackage{amsfonts}
\usepackage{amsmath}
\usepackage{amssymb}
\usepackage{graphicx}%
\begin{document}
\title{Aharonov-Bohm oscillations in the local density of topological surface states}
\author{Zhen-Guo Fu}
\affiliation{State Key Laboratory for Superlattices and Microstructures, Institute of
Semiconductors, Chinese Academy of Sciences, P. O. Box 912, Beijing 100083,
People's Republic of China}
\affiliation{LCP, Institute of Applied Physics and Computational Mathematics, P.O. Box
8009, Beijing 100088, People's Republic of China}
\author{Ping Zhang}
\thanks{zhang\_ping@iapcm.ac.cn}
\affiliation{LCP, Institute of Applied Physics and Computational Mathematics, P.O. Box
8009, Beijing 100088, People's Republic of China}
\affiliation{Center for Applied Physics and Technology, Peking University, Beijing 100871,
People's Republic of China}
\author{Shu-Shen Li}
\thanks{sslee@semi.ac.cn}
\affiliation{State Key Laboratory for Superlattices and Microstructures, Institute of
Semiconductors, Chinese Academy of Sciences, P. O. Box 912, Beijing 100083,
People's Republic of China}

\begin{abstract}
We study Aharonov-Bohm (AB) oscillations in the local density of states (LDOS)
for topological insulator (TI) and conventional metal Au(111) surfaces with
spin-orbit interaction, which can be probed by spin-polarized scanning
tunneling microscopy. We show that the spacial AB oscillatory period in the
total LDOS is a flux quantum $\Phi_{0}\mathtt{=}hc/e$ (weak localization) in
both systems. Remarkably, an analogous weak antilocalization with $\Phi_{0}/2$
periodic spacial AB oscillations in spin components of LDOS for TI surface is
observed, while it is absent in Au(111).

\end{abstract}
\maketitle


Spin-orbit interaction (SOI) effects in semiconductors and metals have been an
active theme in the modern condensed matter physics, but especially, the
topological insulators (TI), which have been detected in series of
two-dimensional \cite{Kane, Bernevig,Konig} and three-dimensional
\cite{Fu,Hsieh1,Chen,Hsieh3,Xia,Moore,Qi,Zhanghj} materials, suggest new
directions for this field since the extraordinarily strong SOI exists in TI.
The helical spin structure of electrons in TI acquire a spin-orbit induced
nontrivial Berry phase of $\pi$ after a $2\pi$ adiabatic rotation along the
Fermi surface. This Berry phase corrects the quantum constructive interference
between a closed trajectory that the electron passes and its time-reversal
counterpart, and thus can give rise to the weak antilocalization (WAL)
signature in the quantum coherent magneto-transport coefficients. Many efforts
have been devoted to observing WAL in the HgTe quantum wells
\cite{Olshanetsky,Tkachov}, Bi$_{2}$Te$_{3}$ \cite{Hasan,He,Ghaemi}, Bi$_{2}%
$Se$_{3}$ \cite{Peng,Bardarson,ZhangY,Chen2,Checkelsky,Liu,Wang}, and other
materials. Very recently, for example, in a transport experiment \cite{Peng}
of Bi$_{2}$Se$_{3}$ nanowires, the Aharonov--Bohm (AB) oscillation with
anomalous period $\Phi_{0}\mathtt{=}hc/e$ (twice the conventional period) of
magnetoconductance was observed, while the WAL induced $\Phi_{0}/2$ period was
absent. The period of these oscillations in conductance is determined by the
doping level and the disorder of the TI nanowire \cite{Bardarson, ZhangY}. In
addition, an energy gap at the Dirac cone opened by magnetic doping in TI
films could also induce crossover from the WAL to the weak localization (WL),
which is tunable by the Fermi energy and the gap \cite{Lu}.

\begin{figure}[ptb]
\begin{center}
\includegraphics[width=0.7\linewidth]{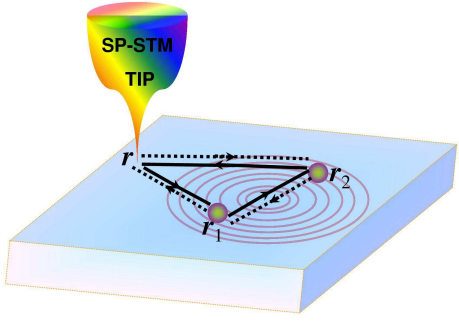}
\end{center}
\caption{ (Color online) Surface electronic interferometer, comprising a spin
polarized STM tip at $\mathbf{r}$ and two impurities at $\mathbf{r}_{1}$ and
$\mathbf{r}_{2}$ separately. The interference contributions in the LDOS is
introduced by the electrons traveling along clockwise and anticlockwise loops
enclosed by the STM tip and two impurities. The applied magnetic field affects
this interference via the AB effect.}%
\label{fig1}%
\end{figure}

In this letter, for exploring the analogous WAL in topological surface state,
we consider an imaginary AB interferometer consisting of a spin-polarized
scanning tunneling microscope (SP-STM) tip \cite{Meier,Bergmann1,Schmaus} and
two identical nonmagnetic impurities apart tens of nanometers on the TI
surface as shown in Fig. \ref{fig1}. When an electron travels along the
clockwise and anticlockwise loops ($\mathbf{r}\mathtt{\rightleftharpoons
}\mathbf{r}_{1}\mathtt{\rightleftharpoons}\mathbf{r}_{2}%
\mathtt{\rightleftharpoons}\mathbf{r}$) enclosing a finite area, the quantum
interference phenomenon occurs. The interference contribution to the local
density of states (LDOS) is affected by an external magnetic field $B$ via the
AB effect arising from the threaded magnetic flux \cite{Cano}. We will show
that, on one hand, the AB oscillatory period of the total LDOS $\Delta
N_{L}\left(  \mathbf{r}\mathtt{,}\omega\mathtt{,}B\right)  $ (the subscription
$L$ represents the loops enclosed by the scattering paths of the surface
electrons) is a flux quantum $\Phi_{0}$, which is an analog of WL phenomenon.
On the other hand, the strong SOI in TI materials can modify the electron AB
interference effect via the chiral spin rotations during the non-collinear
multiple scattering processes, and thereby the analogous WAL effect with
$\Phi_{0}/2$ period in AB oscillations will be observed if the spin-resolved
LDOS [$\Delta N_{L}^{\uparrow}\left(  \mathbf{r}\mathtt{,}\omega
\mathtt{,}B\right)  $ and $\Delta N_{L}^{\downarrow}\left(  \mathbf{r}%
\mathtt{,}\omega\mathtt{,}B\right)  $] are measured. Furthermore, for
comparison, we briefly discuss the AB effect in the LDOS of conventional metal
Au(111) surface with weak SOI, and find that the analogous WAL phenomenon is
absent in Au(111) surface. Our finding may provide a useful method to
characterize the topological surface states. 

We describe the TI surface, on which two nonmagnetic impurities are adsorbed,
by a low-energy effective Dirac Hamiltonian written as
\begin{equation}
H=v\boldsymbol{\sigma}\cdot\left(  \hat{z}\times\mathbf{q}\right)  +V\left(
\mathbf{r}\right)  \label{e1}%
\end{equation}
with the Fermi velocity $v\mathtt{=}333$ (287) meV$\cdot$nm in Bi$_{2}$%
Se$_{3}$ (Bi$_{2}$Te$_{3}$) \cite{Liu3}. $\mathbf{q}\mathtt{=}\left(
q_{x}\mathtt{,}q_{y}\right)  $ denotes the planar momentum operator, and
$\boldsymbol{\sigma}\mathtt{=}\left(  \sigma_{1}\mathtt{,}\sigma_{2}%
\mathtt{,}\sigma_{3}\right)  $ is the Pauli spin matrix. $V$($\mathbf{r}%
$)$\mathtt{=}\sum_{i\mathtt{=}1}^{2}U_{i}\sigma_{0}\delta\left(
\mathbf{r}\mathtt{-}\mathbf{r}_{i}\right)  $ is the potential of two
nonmagnetic impurities located at $\mathbf{r}_{1}\mathtt{=}\left(
\mathtt{-}d/2\mathtt{,}0\right)  $ and $\mathbf{r}_{2}\mathtt{=}\left(
d/2\mathtt{,}0\right)  $ with strength $U_{i}$. $\sigma_{0}$ is the
$2\mathtt{\times}2$ unit matrix.

The features we discuss are expected to be seen in the change of the
real-space LDOS owing to the influence of magnetic flux which passes through
the area enclosed by the two scattering paths shown in Fig. \ref{fig1}. This
quantity can directly reveal the WL or WAL effect in TI via AB oscillatory
periods in LDOS. The real-space Green's function involving the impurities
scattering is given by Dyson equation $G\mathtt{=}G_{0}\mathtt{+}\delta G$,
with
\begin{equation}
\delta G=\int d\mathbf{r}^{\prime\prime}G_{0}\left(  \mathbf{r}-\mathbf{r}%
^{\prime\prime};\omega\right)  V\left(  \mathbf{r}^{\prime\prime}\right)
G\left(  \mathbf{r}^{\prime\prime},\mathbf{r}^{\prime};\omega\right)  .
\label{e4}%
\end{equation}
The bare electron Green's function $G_{0}\mathbf{\ }$for the TI surface states
could be expressed in terms of a complex amplitude multiplied by a
\textquotedblleft complex\textquotedblright\ spin rotation:
\begin{equation}
G_{0}\left(  \mathbf{r,r}^{\prime};\omega\right)  =\frac{\left\vert \omega
^{+}\right\vert }{4v^{2}}D_{+}\hat{R}\left(  \alpha,\beta,\eta\right)  ,
\label{e3}%
\end{equation}
which is useful to gain more physical insight into the transport between
$\mathbf{r}$ and $\mathbf{r}^{\prime}$. Here, $D_{+}\mathtt{=}\sqrt{g_{0}%
^{2}+g_{1}^{2}}$, where $g_{0/1}\mathtt{=}\mathtt{sgn}\left(  \omega
^{+}\right)  Y_{0/1}\left(  \left\vert k\right\vert \rho\right)  \mathtt{\mp
}iJ_{0/1}\left(  \left\vert k\right\vert \rho\right)  $ with $\rho
\mathtt{=}\left\vert \mathbf{r}\mathtt{-}\mathbf{r}^{\prime}\right\vert $,
$k\mathtt{=}k_{F}\left(  1\mathtt{+}\omega/E_{F}\right)  $, $\omega
^{+}\mathtt{=}\omega\mathtt{+}E_{F}$, and $E_{F}\mathtt{=}vk_{F}$ the Fermi
energy of TI. $J_{0/1}$ ($Y_{0/1}$) are the first (second) kind Bessel
functions. $\hat{R}\left(  \alpha,\beta,\eta\right)  \mathtt{=}e^{i\alpha
\sigma_{3}}e^{i\beta\sigma_{1}}e^{i\eta\sigma_{3}}$ is a spin rotation
operator characterized by the three Euler angles $\alpha\mathtt{=}%
\mathtt{-}\frac{\vartheta}{2}\mathtt{+}\frac{\pi}{4}$ with $e^{i\vartheta
}\mathtt{=}(\boldsymbol{\rho}\mathtt{\cdot}\hat{x}\mathtt{+}i\boldsymbol{\rho
}\mathtt{\cdot}\hat{y})/\rho$, $\beta\mathtt{=}\tan^{-1}\mathtt{(}\frac{g_{1}%
}{g_{0}}\mathtt{)}$, and $\eta\mathtt{=}\mathtt{-}\alpha$. Equation (\ref{e3})
is exact under a high-energy cutoff \cite{Liu2009,Biswas2}.

Following the perturbation approach, Eq. (\ref{e4}) can be expanded to any
order in the impurity potential $U_{i}$. Our effort will be concentrated on
the scattering processes of surface electrons with the both impurities, in
which the scattering paths enclose loops. Therefore, taking all this into
account, after a long algebra calculation, we have%
\begin{align}
\delta G_{L}  &  =G_{0}\left(  \mathbf{r}-\mathbf{r}_{1}\right)  W_{1}%
G_{0}\left(  \mathbf{r}_{1}-\mathbf{r}_{2}\right)  T_{2}G_{0}\left(
\mathbf{r}_{2}-\mathbf{r}^{\prime}\right) \nonumber\\
&  +G_{0}\left(  \mathbf{r}-\mathbf{r}_{2}\right)  W_{2}G_{0}\left(
\mathbf{r}_{2}-\mathbf{r}_{1}\right)  T_{1}G_{0}\left(  \mathbf{r}%
_{1}-\mathbf{r}^{\prime}\right)  , \label{loop}%
\end{align}
where
\begin{equation}
W_{1/2}=\frac{T_{1/2}}{\sigma_{0}-T_{1/2}G_{0}\left(  \mathbf{r}%
_{1/2}-\mathbf{r}_{2/1}\right)  T_{2/1}G_{0}\left(  \mathbf{r}_{2/1}%
-\mathbf{r}_{1/2}\right)  } \label{e6}%
\end{equation}
is diagonal with $T\mathtt{-}$matrices $T_{1/2}\mathtt{=}\frac{U_{1/2}%
}{\mathbf{I}\mathtt{-}U_{1/2}G_{0}\left(  \mathbf{0}\mathtt{;}\omega\right)
}$. Equation (\ref{loop}) is a general formula describing the interference
effect from the scattering with both two impurities participated. In the
absence of SOI, the two terms in Eq. (\ref{loop}), i.e., the scattering
amplitudes corresponding to the time-reversal processes, should be equal to
each other and thereby give rise to the constructive quantum correction to the
LDOS in the so-called WL theory. In the presence of a strong SOI, however, the
electrons interference is affected remarkably due to the non-collinear
multiple scattering trajectories that generate nontrivial spin rotations. The
spin rotation leads to a destructive interference in LDOS by a phase change
picked up during the clockwise and anticlockwise scattering processes, which
is the origin of the WAL effect in systems with strong SOI. Explicitly, for
collinear scattering paths between $\mathbf{r}_{1}$ and $\mathbf{r}_{2}$,
there is no net spin rotation since $G_{0}\left(  \mathbf{r}_{1/2}%
\mathtt{-}\mathbf{r}_{2/1}\right)  G_{0}\left(  \mathbf{r}_{2/1}%
\mathtt{-}\mathbf{r}_{1/2}\right)  \mathtt{\varpropto}\sigma_{0}$. Whereas for
non-collinear multiple scattering trajectories, such as the loops shown in
Fig. 1, $G_{0}\left(  \mathbf{r}\mathtt{-}\mathbf{r}_{1/2}\right)
G_{0}\left(  \mathbf{r}_{1/2}\mathtt{-}\mathbf{r}_{2/1}\right)  G_{0}\left(
\mathbf{r}_{2/1}\mathtt{-}\mathbf{r}\right)  $ is not a unit matrix which
implies net spin rotations during scattering processes. As a result, in
contrast to the collinear scattering process, the non-collinear
multiple-impurity scattering on TI surface can induce dramatic modulations in
the LDOS.

Applying a magnetic field tends to destroy the destructive interference and
may bring forth some key signatures of WAL effect, such as the $\Phi_{0}/2$
periodic AB oscillations in the spin components of LDOS (similar to $\Phi
_{0}/2$ oscillations in the magneto-conductance due to WAL). In the presence
of a low magnetic field, the Green's function can be semiclassically
approximated as \cite{Altshuler},%
\begin{equation}
\tilde{G}_{0}\left(  \mathbf{r}-\mathbf{r}^{\prime}\right)  =e^{i\frac{2\pi
}{\Phi_{0}}\int_{\mathbf{r}}^{\mathbf{r}^{\prime}}\mathbf{A}\left(
\mathbf{l}\right)  \cdot d\mathbf{l}}G_{0}\left(  \mathbf{r}-\mathbf{r}%
^{\prime}\right)  , \label{e7}%
\end{equation}
where $\mathbf{A}\mathtt{=}\left(  \mathtt{-}By\mathtt{,}0\mathtt{,}0\right)
$ represents the vector potential. This approximation is exact so long as the
magnetic length is much greater than the Fermi wave length [$l_{B}%
\mathtt{=}\left(  \Phi_{0}/2\pi B\right)  ^{1/2}\mathtt{\gg}\lambda_{F}$]. For
magnetic field $B\mathtt{=}5\mathtt{\sim}10$ T, the corresponding magnetic
length $l_{B}\mathtt{\approx}11.63\mathtt{\sim}8.22$ nm, while the Fermi wave
length $\lambda_{F}\mathtt{=}7.21\mathtt{\sim}5.3$ nm for TI with
$E_{F}\mathtt{=}250\mathtt{\sim}340$ meV \cite{Chen}. Accordingly, the
condition in Eq. (\ref{e7}) for the semiclassical approximation is valid for
these low magnetic field and low energy ranges. Also, the Zeeman splitting is
negligibly small (typically of 1.0 meV at $B\mathtt{=}10$ T for Bi$_{2}%
$Se$_{3}$ film) compared to the strong SOI, and thus is not taken into account
in the following discussion.

The correction of the LDOS due to the magnetic flux is given by%
\begin{equation}
\Delta N_{L}\left(  \mathbf{r},\omega,B\right)  =-\frac{1}{\pi}\Im
\text{Tr}\left[  \delta\tilde{G}_{L}\left(  \mathbf{r},\omega\right)  -\delta
G_{L}\left(  \mathbf{r},\omega\right)  \right]  , \label{e8-1}%
\end{equation}
where $\delta\tilde{G}_{L}$ is calculated from Eq. (\ref{loop}) with
$\tilde{G}_{0}$. For large distances ($\left\vert k\right\vert \rho
\mathtt{\gg}1$), $J_{0/1}\mathtt{\approx}\mathtt{\pm}\sqrt{2/\left(
\pi\left\vert k\right\vert \rho\right)  }\cos\left[  \pi/4\mathtt{\mp
}\left\vert k\right\vert \rho\right]  $ and $Y_{0/1}\mathtt{\approx
}\mathtt{\pm}\sqrt{2/\left(  \pi\left\vert k\right\vert \rho\right)  }%
\sin\left[  \left\vert k\right\vert \rho\mathtt{\mp}\pi/4\right]  $. Combining
Eqs. (\ref{e3}\texttt{-}\ref{e8-1}), we get
\begin{align}
\Delta N_{L}\left(  \mathbf{r},\omega,B\right)   &  =\Delta N_{L}^{\uparrow
}\left(  \mathbf{r},\omega,B\right)  +\Delta N_{L}^{\downarrow}\left(
\mathbf{r},\omega,B\right) \nonumber\\
&  \approx C\Im\lbrack\tilde{t}\left(  \omega\right)  e^{i\left[  \chi\left(
r\right)  -\pi/4\right]  }]\left[  \cos\left(  \frac{2\pi\Phi}{\Phi_{0}%
}\right)  -1\right]  , \label{e9}%
\end{align}
where $C\mathtt{\propto}k^{-3/2}\sqrt{\frac{2}{d\rho_{1}\rho_{2}}}%
\cos\mathtt{(}\frac{\vartheta_{2}}{2}\mathtt{)}\sin\mathtt{(}\frac
{\vartheta_{1}}{2}\mathtt{)}\sin\mathtt{(}\frac{\vartheta_{1}\mathtt{-}%
\vartheta_{2}}{2}\mathtt{)}$ with $e^{i\vartheta_{1/2}}\mathtt{=}%
(\boldsymbol{\rho}_{1/2}\mathtt{\cdot}\hat{x}\mathtt{+}i\boldsymbol{\rho
}_{1/2}\mathtt{\cdot}\hat{y})/\rho_{1/2}$ and $\boldsymbol{\rho}%
_{1/2}\mathtt{=}\mathbf{r}\mathtt{-}\mathbf{r}_{1/2}$, $\chi\left(  r\right)
\mathtt{=}k\left(  d\mathtt{+}\rho_{1}\mathtt{+}\rho_{2}\right)  $. $\tilde
{t}\left(  \omega\right)  \mathtt{=}\frac{t^{2}\left(  \omega\right)
}{1\mathtt{-}t^{2}\left(  \omega\right)  \left[  g_{0}^{2}\left(  d\right)
\mathtt{-}g_{1}^{2}\left(  d\right)  \right]  }$ with $t\left(  \omega\right)
\mathtt{=}\frac{U}{1\mathtt{-}U\mathtt{\int}\frac{d^{2}q}{\left(  2\pi\right)
^{2}}\frac{i\omega}{\left(  i\omega\right)  ^{2}\mathtt{-}\left(  vq\right)
^{2}}}$ is the nonzero element of $T\mathtt{-}$matrix. One can notice that
$\left[  g_{0}^{2}\left(  d\right)  \mathtt{-}g_{1}^{2}\left(  d\right)
\right]  $ vanishes within the above asymptotic representation of $J_{0/1}$
and $Y_{0/1}$.

In the present setup, we focus solely on the (spin-resolved) LDOS at
$\mathbf{r}\mathtt{=}(x,y)$, which is probed by the STM tip with the same
plane coordinates $(x,y)$. Since the STM tip also dually plays as a scatter
for the electron's loop motion, thus, in the presence of a fixed magnetic
field $B$, the AB oscillation varies with the tip position along the $y$
direction. Obviously, the AB oscillation signatures result from $\cos
\mathtt{(}\frac{2\pi\Phi}{\Phi_{0}}\mathtt{)}\mathtt{-}1\mathtt{=}0$ in Eq.
(\ref{e9}), and the spacial oscillation period is $d_{0}\mathtt{=}\frac
{2\Phi_{0}}{Bd}$ for fixed $B$ and fixed impurities configuration,
corresponding to a $\Phi_{0}$ period in the scale of flux, which gives rise to
WL effect in the total LDOS. To observe a complete AB oscillation period in
$\Delta N_{L}\left(  \mathbf{r}\mathtt{,}\omega\mathtt{,}B\right)  $, the
magnetic field should satisfy the relation
\begin{equation}
B_{2\pi}\sim2\Phi_{0}/yd. \label{e12}%
\end{equation}

Now let us analyze the spin-up and spin-down components of LDOS. The
extraordinary SOI in TI affects the surface electron interference due to the
non-collinear multiple scattering trajectories that generate nontrivial spin
rotations, which can be easily obtained by the rotation operator $\hat
{R}\left(  \alpha,\beta,\eta\right)  $ in the Green's function $G_{0}$.
Explicitly, the AB oscillations of spin components of LDOS are given by
\begin{align}
\Delta N_{L}^{\uparrow/\downarrow}\left(  \mathbf{r},\omega,B\right)    &
\approx\mp C_{s}\operatorname{Im}\left[  \tilde{t}(\omega)e^{i\left(
\chi\left(  r\right)  -\frac{\pi}{4}\right)  }\right]  \label{e10}\\
& \times\left[  \sin\left(  \frac{2\pi\Phi}{\Phi_{0}}\mp\phi\right)  \pm
\sin\phi\right]  ,\nonumber
\end{align}
where $\phi\mathtt{=}\frac{\vartheta_{1}\mathtt{-}\vartheta_{2}}{2}%
\mathtt{=}\frac{1}{2}\mathtt{[}\tan^{\mathtt{-}1}(\frac{y}{x\mathtt{+}%
d/2}\mathtt{)}\mathtt{-}\tan^{\mathtt{-}1}\mathtt{(}\frac{y}{x\mathtt{-}%
d/2}\mathtt{)]}$ and $C_{s}$=$\frac{C}{2}\sin\phi$. As clearly seen from this
equation, comparing to the total LDOS, there occur in the spin-resolved LDOS
additional strong SOI induced quantum interference signature. The above
equation should be reasonable because of the destructive interference brought
about by the spin rotation during clockwise and anticlockwise scattering
processes as well as by the magnetic field. This strong spin interference
effect deviates the real-space AB oscillations from $d_{0}$ period when a
SP-STM tip scans on the TI surface in the presence of a fixed $B$, which is
our concentration in the present paper. We can easily find that the spacial AB
oscillation period of $\Delta N_{L}^{\uparrow/\downarrow}$ is determined by
the factor $F^{\uparrow/\downarrow}\mathtt{=}\sin(\frac{2\pi\Phi}{\Phi_{0}%
}\mathtt{\mp}\phi)\mathtt{\pm}\sin\phi$. Taking $\Delta N_{L}^{\uparrow}$ as
an example (Similar analysis can be done on $\Delta N_{L}^{\downarrow}$) to
find the AB period, we consider the roots of $F^{\uparrow}\mathtt{=}0$ as an
equation of $y$, which can be rewritten as $F^{\uparrow}\mathtt{=}\sin
\phi\lbrack\cos(\frac{2\pi\Phi}{\Phi_{0}})\mathtt{-}1]\mathtt{-}\cos\phi
\sin(\frac{2\pi\Phi}{\Phi_{0}})\mathtt{=}0$. There are two cases that result
in $F^{\uparrow}\mathtt{=}0$: (i) First, it is easy to get that $y_{1}%
\mathtt{=}\frac{2n\Phi_{0}}{Bd}$ $\left(  n\mathtt{=}0,\mathtt{\pm
}1,\mathtt{\cdots}\right)  $ are the roots of $F^{\uparrow}\mathtt{=}0$; (ii)
Second, expanding $F^{\uparrow}$ at point $y_{2}\mathtt{=}\frac{\left(
2n\mathtt{+}1\right)  \Phi_{0}}{Bd}$, a simple equation for $y$ can be
obtained $M(y\mathtt{-}\frac{\left(  2n\mathtt{+}1\right)  \Phi_{0}}%
{Bd})\mathtt{-}P\mathtt{=}0$, which gives out other asymptotic roots
$y\mathtt{=}\frac{\left(  2n\mathtt{+}1\right)  \Phi_{0}}{Bd}\mathtt{+}%
\frac{P}{M}$ ($M$ and $P$ are the expanding coefficients). These roots lie
near $y_{2}$ and become better for the limit of small $B$ [see following Figs.
\ref{fig2}(b) and \ref{fig2}(c)]. Combining with case (i), the AB oscillation
signals for $\Delta N_{L}^{\uparrow}$ occur at $\mathtt{\sim}\frac{n\Phi_{0}%
}{Bd}$ with a spacial period of $\frac{\Phi_{0}}{Bd}\mathtt{=}\frac{d_{0}}{2}$
(i.e., $\frac{\Phi_{0}}{2}$ in the scale of flux), the half of $\Delta N_{L}$.
This half period could be understood as an analog of WAL effect in the
spin-polarized LDOS since strong SOI in TI can result in the two-dimensional
WAL effect in the magneto-conductance, which can be represented as the AB
oscillation phenomenon with period of $\frac{\Phi_{0}}{2}$.

Moreover, the interference signature of LDOS decays by a factor of $\left[
d\rho_{1}\rho_{2}\right]  ^{-1/2}$ in the asymptotic representations, Eqs.
(\ref{e9}) and (\ref{e10}). Actually, however, dephasing processes have been
observed in transport investigations in Bi$_{2}$Se$_{3}$ and Bi$_{2}$Te$_{3}$
films \cite{Checkelsky, Wang, Liu, He} as well as in AB-effect studies of
Bi$_{2}$Se$_{3}$ nanowires \cite{Peng,Bardarson,ZhangY}. The phase coherence
length $l_{\phi}$ of Bi$_{2}$Se$_{3}$ and Bi$_{2}$Te$_{3}$ can be as large as
hundreds of nanometers, which is tens times of the Fermi wave length. The
characteristic distance in our setup must be much smaller than the phase
coherence length ($d\mathtt{\leq}l_{\phi}$), so that we choose $d\mathtt{=}%
20$nm$\mathtt{\ll}l_{\phi}$ without taking into account the dephasing
processes in the following numerical calculations. \begin{figure}[ptb]
\begin{center}
\includegraphics[width=1.\linewidth]{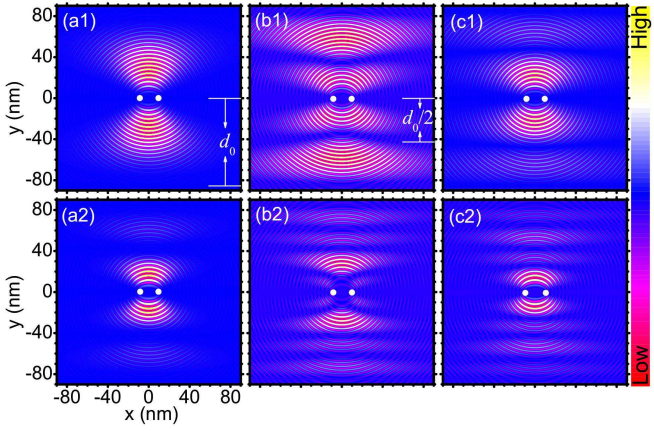}
\end{center}
\caption{ (Color online) Simulations of the AB oscillations of the electronic
LDOS in Bi$_{2}$Te$_{3}$(111) surface with fixed magnetic field $B\mathtt{=}5$
T (upper panels) and $B\mathtt{=}10$ T (lower panels). (a) the total LDOS
patterns; (b) the spin-up component; (c) the spin-down component. The
parameters are chosen as $v\mathtt{=}287$ meV$\cdot$nm, $E_{F}\mathtt{=}250$
meV ($\lambda_{F}\mathtt{=}7.21$ nm) and $d\mathtt{=}20$ nm. The white dots
denote the positions of adatoms. The periodic horizonal blue strips in
patterns are signature of AB oscillations in LDOS.}%
\label{fig2}%
\end{figure}

To supplement the above analytical results, typically, we present our
numerically calculated data in Fig. \ref{fig2}, where the upper (lower) panels
correspond to $B\mathtt{=}5$ T ($10$ T). The blue horizontal strips are the AB
oscillation signals in the real-space LDOS, while the oscillatory ellipse
features are the interference signals arising from the contributions of
$N_{L}\left(  \mathbf{r}\mathtt{,}\omega\mathtt{,}B\mathtt{=}0\right)  $. It
is obvious that in the case of $B\mathtt{=}5$ T ($10$ T), the interstrip
distance is $d_{0}\mathtt{\approx}85$ nm ($42$ nm) in the total LDOS as shown
in Fig. \ref{fig2}(a), corresponding to the $\Phi_{0}$ period of AB
oscillations. The numerical data in Fig. \ref{fig2}(a) are exact while Eq.
(\ref{e9}) is approximate.

However, differing from the $\Phi_{0}$ periodic AB oscillations in the total
LDOS, Fig. \ref{fig2}(b) and \ref{fig2}(c) show that the interstrip distance
in the spin-resolved LDOS is $\frac{d_{0}}{2}$, corresponding a $\frac
{\Phi_{0}}{2}$ period. This analogous phenomenon of WAL with a period of
$\frac{\Phi_{0}}{2}$ in $\Delta N_{L}^{\uparrow/\downarrow}\left(
\mathbf{r},\omega,B\right)  $ found from the numerical calculation are accord
well with our analytical result given by Eq. (\ref{e10}). As addressed above,
the present analogous WAL phenomenon can be understood via the Berry phase.
Explicitly, the quantum phase difference between the closed loops in opposite
directions in our setup corresponds to the Berry phase associated with spin
rotation by $2\pi$, which is given by $\Delta\varphi\mathtt{=}\mathtt{-}%
i\int_{\theta_{0}}^{\theta_{0}+2\pi}d\theta\left\langle \psi_{q}\right\vert
\frac{\partial}{\partial\theta}\left\vert \psi_{q}\right\rangle \mathtt{=}%
\mathtt{\pm}\pi$. Here, $\left\vert \psi_{q}\right\rangle \mathtt{=}\frac
{1}{\sqrt{2}}\left(  \pm ie^{i\theta\left(  q\right)  },1\right)  ^{T}$ are
the eigenstates for the free Hamiltonian of TI surface with $\tan\left[
\theta\left(  q\right)  \right]  \mathtt{=}\frac{q_{y}}{q_{x}}$. To
experimentally verify our predicted AB interference strips with $\frac
{\Phi_{0}}{2}$peroid in the spin-resolved LDOS shown in Fig. \ref{fig2}(b) and
\ref{fig2}(c), the SP-STM tip and sufficiently strong scattering potentials
are required, which we believe are achievable in current experimental
capabilities. So we hope the present prediction can be directly observed by
STM \textit{in situ} measurement instead of the complicated low-temperature
transport measurement.

\begin{figure}[ptb]
\begin{center}
\includegraphics[width=1.\linewidth]{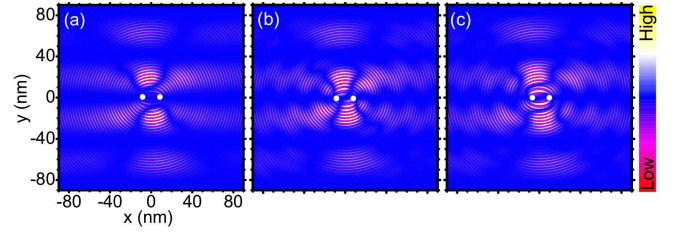}
\end{center}
\caption{ (Color online) Simulated STM maps of the AB oscillations of the LDOS
for two identical impurities 20 nm apart on the Au(111) surface with a
SOI\ constant $\gamma\mathtt{=}40$ meV$\cdot$nm under $B\mathtt{=}10$ T. (a)
the total LDOS pattern, (b) and (c) the spin-up and spin-down components,
separately. The effective electron mass $m^{\ast}\mathtt{=}0.26m_{0}$ and the
Fermi wave length $\lambda_{F}\mathtt{=}3.74$ nm have been chosen.}%
\label{fig3}%
\end{figure}

For comparison, we also calculate the AB oscillations on the conventional
metal surface with weak but observable SOI. We choose Au(111) as an example,
in which the Rashba SOI is $\mathtt{\sim}40$ meV$\cdot$nm and the Fermi wave
length is $\mathtt{\sim}3.74$ nm \cite{Walls}. The consequent calculated
results are shown in Fig. \ref{fig3} for $B\mathtt{=}10$ T. From Fig.
\ref{fig3} the SOI influence on the electron interference in Au(111) can be
summarized as follows: (i) The AB interference strips in the spin-resolved
LDOS display oscillatory behavior along $x$ axis, which is caused by the spin
rotations induced by non-collinear multiple scattering trajectories; (ii) The
SOI destroys the elliptic features in the LDOS maps even if there is no
applied magnetic field, see the \textit{longitudinal} extending blue strips in
Fig. \ref{fig3}. However, the AB oscillation period in the LDOS pattern is
also $\Phi_{0}$ (the interstrip distance is $\mathtt{\sim}42$ nm); (iii)
Especially, there is no $\Phi_{0}/2$ period in the spin-resolved LDOS in Fig.
\ref{fig3} because there is zero net Berry phase and no topological chirality
in the Shockley surface state on Au(111), which is totally different from the
case of TI surface. Analytically, the unperturbed spatial Green's function for
the conventional metal surface with weak Rashba SOI described by the
Hamiltonian $H_{0}^{(c)}$=$\left(  \hbar q\right)  ^{2}/2m^{\ast}%
\mathtt{-}\left(  \gamma/\hbar\right)  (\boldsymbol{\sigma}\mathtt{\times
}\mathbf{q})\mathtt{\cdot}\hat{z}$ is asymptotically expressed as
$G_{0}\left(  \mathbf{r},\mathbf{r}^{\prime},\omega\right)  \mathtt{\approx
}\tau\left[  f_{\mu}\left(  \rho,\omega\right)  \mathtt{+}f_{\nu}\left(
\rho,\omega\right)  \left(  \boldsymbol{\sigma}\mathtt{\cdot}\hat{\rho
}\right)  \right]  $,\qquad where $\tau\mathtt{=}\mathtt{-}\frac{m^{\ast
}(1\mathtt{+}i)}{2(k_{1}\mathtt{+}k_{2})\hbar^{2}\sqrt{\pi}}$,
$\boldsymbol{\rho}\mathtt{=}\mathbf{r}\mathtt{-}\mathbf{r}^{\prime}$, and
$f_{\mu/\nu}\left(  \rho,\omega\right)  \mathtt{\approx}\sqrt{\frac{k_{1}%
}{\rho}}e^{ik_{1}\rho}\mathtt{\pm}\sqrt{\frac{k_{2}}{\rho}}e^{ik_{2}\rho}$
with $k_{1/2}\mathtt{=}\mathtt{\pm}k_{so}\mathtt{+}\sqrt{k_{so}^{2}%
\mathtt{+}\frac{2m^{\ast}\left(  \omega\mathtt{+}E_{F}\right)  }{\hbar^{2}}}$
and $k_{so}\mathtt{=}m^{\ast}\gamma/\hbar^{2}$. After a tedious derivation, we
find that for the conventional metal surface with the same interferometer
setup shown in Fig. \ref{fig1}, the correction in the spin-resolved LDOS due
to the magnetic flux can be approximated by%
\begin{align}
\Delta N_{L}^{\uparrow/\downarrow}\left(  \mathbf{r},\omega,B\right)    &
\approx-2\operatorname{Im}\left[  \tilde{t}\left(  \omega\right)  \tau
^{3}f_{\mu}\left(  \rho_{1}\right)  f_{\mu}\left(  d\right)  f_{\mu}\left(
\rho_{2}\right)  \right]  \nonumber\\
& \times\left[  \cos\left(  2\pi\Phi/\Phi_{0}\right)  -1\right]  ,
\end{align}
from which the absence of $\frac{\Phi_{0}}{2}$ AB oscillatory period becomes
obvious. 

In summary, we have performed a semiclassical analysis of the SP-STM probed AB
oscillations in the LDOS induced by two impurities on a TI surface as well as
on a conventional metal surface with spin splitting. We have found that the
total LDOS in both systems present WL phenomenon with an oscillatory period
$\Phi_{0}$ in the AB oscillations. Remarkably, the analogous WAL signified by
a $\frac{\Phi_{0}}{2}$ oscillation period has been found in the spin-resolved
LDOS in the TI system, while it was absent in the conventional metal surfaces.
This phenomenon, which can be observed in the SP-STM experiments, may provide
an important signature for the existence of the topological surface states and
provide a useful criterion to distinguish the TI surface from other
two-dimensional systems.

This work was supported by NSFC under Grants No. 90921003, No. 60776063, and
No. 60821061, and by the National Basic Research Program of China (973
Program) under Grants No. 2009CB929103 and No. G2009CB929300.

\end{document}